\documentclass[conference]{IEEEtran}
\IEEEoverridecommandlockouts
\usepackage{cite}
\usepackage{amsmath,amssymb,amsfonts}
\usepackage{algorithmic}
\usepackage{graphicx}
\usepackage{textcomp}
\usepackage[table]{xcolor}
\usepackage{xcolor}
\usepackage{siunitx}
\usepackage{makecell}  
\usepackage{booktabs}
\usepackage{tikz,xcolor,hyperref}
\usepackage{array}
\usepackage{url}
\usepackage{subcaption}

\def\BibTeX{{\rm B\kern-.05em{\sc i\kern-.025em b}\kern-.08em
    T\kern-.1667em\lower.7ex\hbox{E}\kern-.125emX}}
\begin{document}
\title{PhishGuard: A Convolutional Neural Network-Based Model for Detecting Phishing URLs with Explainability Analysis\\
}


\author{


\begin{tabular}[t]{@{}c@{}}
Md Robiul Islam \\
\textit{Dept. of Computer Science and} \\
\textit{Engineering} \\
\textit{Uttara University} \\
Dhaka, Bangladesh \\
robiul.cse.uu@gmail.com
\end{tabular}

\and


\begin{tabular}[t]{@{}c@{}}
Md Mahamodul Islam \\
\textit{Dept. of Informatics} \\
\textit{University of Oslo} \\
Oslo, Norway \\
mdmahamodul1998@gmail.com
\end{tabular}

\and


\begin{tabular}[t]{@{}c@{}}
Mst. Suraiya Afrin \\
\textit{Dept. of Building Engineering and} \\
\textit{Construction Management} \\
\textit{Khulna University of} \\
\textit{Engineering and Technology (KUET)} \\
Khulna, Bangladesh \\
abontyefu@gmail.com
\end{tabular}

\and


\begin{tabular}[t]{@{}c@{}}
Anika Antara \\
\textit{Dept. of Electrical and} \\
\textit{Electronics Engineering} \\
\textit{BRAC University} \\
Dhaka, Bangladesh \\
anikaantara1000@gmail.com
\end{tabular}

\and


\begin{tabular}[t]{@{}c@{}}
Nujhat Tabassum  \\
\textit{Dept. of Electrical and} \\
\textit{Computer Engineering} \\
\textit{North South University} \\
Dhaka, Bangladesh \\
nujhat1tabassum2@gmail.com
\end{tabular}

\and


\begin{tabular}[t]{@{}c@{}}
Al Amin \\
\textit{Dept. of Computer Science} \\
\textit{American International} \\
\textit{University Bangladesh (AIUB)} \\
Dhaka, Bangladesh \\
alaminbhuyan321@gmail.com
\end{tabular}

}



\IEEEoverridecommandlockouts
\IEEEpubid{\makebox[\columnwidth]{978-1-5386-5541-2/18/\$31.00~\copyright2024 IEEE \hfill}
\hspace{\columnsep}\makebox[\columnwidth]{ }}
\maketitle

\begin{abstract}
Cybersecurity is one of the global issues because of the extensive dependence on cyber systems of individuals, industries, and organizations. Among the cyber attacks, phishing is increasing tremendously and affecting the global economy. Therefore, this phenomenon highlights the vital need for enhancing user awareness and robust support at both individual and organizational levels. Phishing URL identification is the best way to address the problem. Various machine learning and deep learning methods have been proposed to automate the detection of phishing URLs. However, these approaches often need more convincing accuracy and rely on datasets consisting of limited samples. Furthermore, these black-box intelligent models' decision to detect Suspicious URLs needs proper explanation to understand the features affecting the output. To address the issues, we propose a \textit{1D Convolutional Neural Network (CNN)} and trained the model with extensive features and a substantial amount of data. The proposed model outperforms existing works by attaining an accuracy of 99.85\%. Additionally, our explainability analysis highlights certain features that significantly contribute to identifying the phishing URL.
\end{abstract}

\section{Introduction}\label{sec:1}
The increasing digitization of our daily lives reflects the rapid evolution of global networks and communication tools. Cyberattacks gain in this anonymous playground, leaving everyone vulnerable to data breaches, malware infections, and sophisticated scams. Phishing scams are designed to exploit human vulnerabilities, making it inherently difficult to completely prevent users from falling for them, regardless of their caution or experience \cite{one}. Skilled phishers recognize that personality plays a role in online behaviour and capitalize on that to deceive even the most experienced users who might otherwise be alert \cite{two}. Ordinary online users are the new battleground for cybercrime. These targeted attacks steal precious personal data and money, draining billions from individuals every year \cite{three}. Exploiting these deceitful URLs, phishers aim to harvest sensitive and personal data from victims, including financial details, personal information, usernames, passwords, and more \cite{four}. If users mistakenly access such fraudulent sites, assuming they are legitimate, they may unwittingly provide their sensitive information without suspicion. This is because the deceptive web page appears identical to the original one. In a correlated investigation concerning user encounters with phishing attacks \cite{five}, computer users fall for phishing due to a limited understanding of URLs, the lack of knowledge of identifying trustworthy websites, haste or accidental clicks, difficulty in assessing the true nature of a link and the challenge of spotting phishing scams.

In this paper, we focus on creating an intelligent real-time phishing web page detection system. For the establishment of a robust model, we have created a large volume of training data and extracted features. Finally, we have proposed an explainable 1D Convolutional Neural Network (CNN). Our main contributions to the research work are as follows - 
\begin{itemize}
     \item We proposed a fine-tuned 1D Convolutional Neural Network (CNN) based phishing detection model, which outperforms other benchmark models. 
    \item We have done extensive analysis from different perspectives to evaluate the performance of the model. Creating a large dataset and extracting numerous features are the key contributions that positively impact the model's performance.  
    \item To comprehend the internal mechanisms of our proposed model, we have analyzed its explainability. It will help users to be cautious about the substantial features contributing to phishing URLs.   
\end{itemize}
The rest of the paper is organized as follows: Section \ref{relatedWork} refers to related works about phishing detection. In the next Section \ref{dataset}, we discussed the dataset creation and pre-processing. Furthermore, in Section \ref{methodology}, we have discussed the methodology of our proposed model. Next, we have discussed the analysis of the result in Section \ref{Evaluation}. In Section \ref{conclusion}, we conclude our work.

\section{Related Work}\label{relatedWork}
Numerous researchers have investigated the outcomes associated with phishing websites. Our method incorporates fundamental principles derived from previous research findings. Previous URL-based phishing detection initiatives significantly influenced our current approach. 

Aljofey et al. \cite{fifteen} proposed a method to identify phishing based on a specific publication's URL; the researchers employed multiple algorithms. They analyzed diverse data points using different deep-learning models and layered architectures to compare the outputs. The initial method examines multiple URL characteristics, furthermore, assesses the website's credibility by analyzing its affiliations and administration, and the third evaluates its visual presentation. Yao et al. \cite{sixteen} suggested a revolutionary strategy for identifying phishing websites, relying solely on URL analysis, which has been hailed for its remarkable accuracy and effectiveness. One strategy involves initiating by eliminating superficial URL property. However, in [ \cite{seventeen}, \cite{sixteen}], the researchers create reliable and precise complex characteristics of URLs, leveraging basic features to assess URL authenticity. Aggregating outcomes from all segments determine the overall effectiveness of this technology. Korkmaz et al. \cite{eighteen} introduced a technique to detect phishing webpages using Hypertext Markup Language (HTML), Tag Distribution Language (HTDL), and URL properties. They developed compact HTDL and URL properties, enabling the creation of HTDL string-embedding functions without external dependencies, facilitating integration into a legitimate recognition application. The effectiveness of the method was confirmed through testing on a substantial dataset comprising more than 30,000 HTDL and URL attributes. The authors claim their system reached 98.24\% accuracy, with a 3.99\% True Positive rate and a 1.74\% False Negative rate \cite{eighteen}. To identify a brand-new phishing scheme targeting vulnerabilities. Stokes et al. \cite{nineteen} proposed a exceptional strategy for intelligent phishing detection considering the website url properties. The researchers utilized resource identification principles and sequencing strategies consistent with their framework's usual practices. As per the researcher's findings, the proposed method effectively identifies phishing attempts and zero-day attacks, with a True Positive rate of 95.38\%. Earlier research used how website text is organized to make tools for spotting phishing.

Yerima et al. \cite{tweenty} explored the effectiveness of the long short-term memory (LSTM) classifier, contrasting it with a Random Forest (RF) classifier-based approach through a scientific investigation of web addresses involving fourteen criteria. Initially, researchers constructed an LSTM model that treats a hyperlink as a textual sequence, discerning its legitimacy. Despite its feature creation not necessitating specialized expertise, users found that the LSTM algorithm surpasses on average, the accuracy rates of the RF classifier. Even if there's no preparation or arrangement of features, their approach acquire 97.4\% accuracy \cite{tweenty}. Their study primarily focused on the textual features of webpages, neglecting to incorporate additional elements like frame characteristics and website images, which could potentially improve the model's effectiveness. Selamat et al. \cite{tweentyone}  employed two sets of URL data to detect phishing, employing logistic regression alongside CNN and CNN-LSTM models. These datasets were compiled from diverse sources, including malware domains, lists of malware domains, and phishing domains sourced from OpenPhish and PhishTank. The dataset contains over 70,000 URLs for training purposes and over 60,000 URLs for testing. They utilized this dataset to train models for detecting phishing URLs using both CNN-LSTM and CNN. The LSTM method was chosen because it considers the raw web address as input data. In their experiment, the CNN- LSTM architecture surpassed the alternative model, achieving an accuracy rate of approximately 97\% in categorizing URLs \cite{tweentytwo}. On the contrary, the proposed method relies solely on text-based features and could benefit from augmenting with additional attributes and fine-tuning variables to enhance accuracy. Hence, the limitations observed in earlier research studies formed the basis for our introduced IPDS. Janet et al. \cite{tweentythree}  utilized deep learning techniques, and they aimed to identify phishing activities. Employing classification methods like Support Vector Machine (SVM), ANN, and RF, they found that the RF algorithm demonstrated strong performance. Rishi Kotak \cite{tweentyfour} identified phishing attacks utilizing different deep learning techniques. Finally, Rabab et al. \cite{tweentyfive} tested DL models for spotting phishing attempts, and the results showed that random forests performed exceptionally well. 

Reviewing the literature on Phishing Detection in Modern Security focusing on URLs exposed various potential research opening. One such gap involves the absence of standardization in identifying phishing URLs, complicating result comparisons for researchers and the implementation of robust phishing spotting systems by organizations. Another gap pertains to limited research on real-world data; most studies rely on synthetic or artificially generated datasets, potentially overlooking the complexity of actual phishing attacks. Furthermore, many studies solely concentrate on the technical aspects of phishing detection, disregarding user interaction with phishing URLs. It's important to understand user behavior and decision-making processes to create successful phishing detection strategies. Lastly, there needs to be more focus on emerging threats; studies tend to target known phishing attack types without taking into account new methods of attack. Our proposed model aims to bridge these gaps by using a 1D CNN-based approach to precisely classify legitimate websites from phishing ones. Its effectiveness has been assessed using the PhishTank dataset. We have also analyzed the explainability of the model.
\section{Dataset creation and pre-processing}\label{dataset}
A well-crafted dataset, which may be the secondary or primary data, fuels any Machine Learning (ML) model. We must care about the data; it should be representative and free of null, garbage, and missing values. Otherwise, we can't think of having a good ML model. If we input garbage data, it will produce a garbage model \cite{kilkenny2018data}. So, collecting or creating a training dataset is our model's first and critical stage. We will vividly discuss the entire process in the following subsections. 

\subsection{Dataset collection}
Throughout our work, we used 
\textbf{PhishTank} dataset \cite{PhishTank}. The dataset consists of URLs of websites that people report as scams. This dataset has information about these sites, like their \textit{web addresses, when they were reported, whether they're still active or not, etc}. There are almost 38,000 instances and 8 features. We can easily get it through an API or as a CSV file format. This data helps researchers make better tools to fight against online scams. It gets updated daily, so using the newest version is important. 


\noindent The legitimate URLs are obtained from the \textit{open datasets of the University of New Brunswick}. This dataset consists of 35000 authentic URLs \cite{URL2016D41}. For demonstration purposes, we have shown some sample data in Table \ref{table:legitimateDataTable}. All of the URLs of the dataset are legitimate web addresses. We used these two datasets to create our training dataset. 

\begin{table}[!ht]
  \centering
  \caption{\centering Legitimate Dataset Sample Data}
  \label{table:legitimateDataTable}
  \begin{tabular}{cc}
    \toprule
    \textbf{Index} & \textbf{URLs} \\
    \midrule
    1 & \url{https://lastpass.com/signup2.php?ac=1&from_uri...} \\
    2 & \url{http://persianblog.ir/wEPDwUKMTgwNjcyNjU0MA9kF...} \\
    3 & \url{https://twitter.com/share?text=%D0%9D%D0%B0+%D...} \\
    4 & \url{https://asana.com/guide/videos/%22//fast.wisti...} \\
    \bottomrule
\end{tabular}
\end{table}

\subsection{Data Preprocessing and Feature extraction} This is the most essential part of our model, and we put a lot of effort into preprocessing the data and extracting the features. From both of the datasets (phishing and non-phishing), we only select the features \textit{'URLs' and 'Labels'}. To extract the features, we used different custom functions to create each function based on specific criteria. Those functions extract the features based on different criteria like \textit{IP Address},\textit{Abnormal URL}, and so on. We used regular expressions and some string functions to extract the features from the URLs. We set 21 different criteria to have 21 features.  Finally, we merge two datasets to create our final training dataset. 
\section{Methodology}\label{methodology}
\begin{figure}[!h]
    \centering
    \includegraphics[height=0.4\textheight]{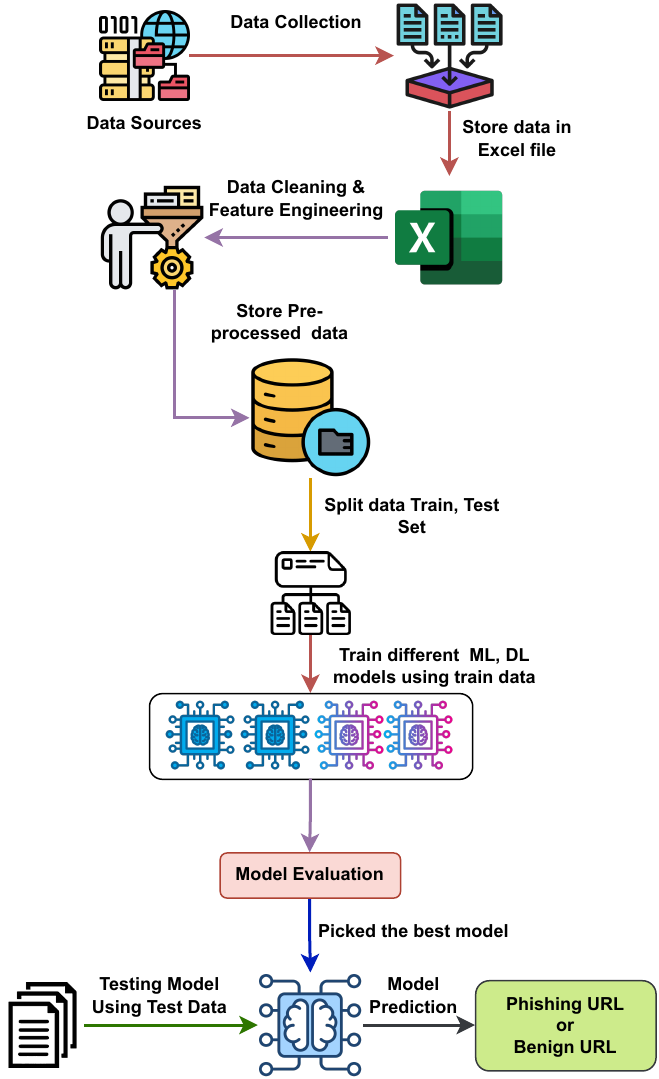}
    \caption{\centering Proposed Methodology Workflow}
    \label{fig:methodologyImg}
\end{figure}

This section describes our proposed CNN-based phishing URL detection method. The suggested method consists of a few phases that work jointly to accomplish high accuracy in recognizing fraudulent websites. Figure \ref{fig:methodologyImg} shows the overall approach of our work.
The initial step involves dataset preparation and preprocessing, which is clearly described in the previous section \ref{dataset}. The last step is to develop a 1D CNN model to detect the phishing URLs. Though CNN is a well-known deep learning-based image analysis model, we have customized the model for URL detection. Lastly, we also outline the attributes of the various layers within the CNN network, which allows us to make sense of the trained model's inner working principles to identify the important features for detecting phishing URLs.

\subsection{Proposed CNN Model}
After using a couple of machine learning algorithms, we have seen that the deep learning algorithm performs far better than the machine learning algorithm. We have chosen CNN as it produces the best accuracy. We have fine-tuned the CNN model according to the following manner. Table \ref{table:summaryofCNNModel} presents a summary of the proposed Convolutional Neural Network (CNN) model, detailing the constituent layers of the model alongside the corresponding count of parameters associated with each layer.



\begin{table}[!hb]
  \centering
  \caption{\centering Summary of the proposed CNN model}
  \label{table:summaryofCNNModel}
  \begin{tabular}{llS[table-format=4.0]}
    \toprule
    \multicolumn{3}{c}{\textbf{Model: Sequential}} \\
    \midrule
    \textbf{Layer (type)} & \textbf{Output Shape} & \textbf{Param} \\
    \midrule
    Conv1D & (None, 19, 32) & 128 \\
    MaxPooling1D & (None, 9, 32) & 0 \\
    Conv1D & (None, 7, 64) & 6208 \\
    GlobalAveragePooling1D & (None, 64) & 0 \\
    Dense & (None, 64) & 4160 \\
    Batch\_Normalization & (None, 64) & 256 \\
    Dense\_1 & (None, 1) & 65 \\
    \midrule
    \multicolumn{3}{c}{Total Params: 10817} \\
    \multicolumn{3}{c}{Trainable Params: 10689} \\
    \multicolumn{3}{c}{Non-trainable Params: 128} \\
    \bottomrule
  \end{tabular}
\end{table}

\section{Evaluation and Explainability Analysis}\label{Evaluation}
The evaluation matrix, also known as performance metrics, is an essential aspect of machine learning (ML) and deep learning (DL). It refers to the set of metrics used to measure the effectiveness of an ML and DL model. These metrics help to determine how well a model can make accurate predictions or classifications on unseen data. Several evaluation metrics such as \textit{\textbf{accuracy}}, \textit{\textbf{precision}}, \textit{\textbf{recall}}, and \textit{\textbf{F1-score}}  used to evaluate our machine and deep learning model \cite{evaluation}. 
\newline
We trained multiple machine learning models and selected the best model based on the performance. For the purpose of execution, we utilized a personal computer for this task. The details of the computer's specifications are summarized in Table \ref{table:system_configuration}. In the following subsections, we have discussed all of our experimental results sequentially. 
\begin{table}[!ht]
  \centering
  \caption{\centering System Configuration.}
  \label{table:system_configuration}
  \begin{tabular}{cc}
    \toprule
    \textbf{Component} & \textbf{Specification} \\
    \midrule
    OS & Windows 11 (64-bit) \\
    Processor & AMD Ryzen 5 5600G with Radeon Graphics \\
    RAM & 16GB \\
    Storage & HP SSD EX900 500GB \\
    \bottomrule
\end{tabular}
\end{table}

\subsection{Accuracy and Loss for Training and Testing Phase}
\begin{figure}[!ht]
  \centering
  \begin{subfigure}[b]{0.47\textwidth}
    \includegraphics[width=\textwidth]{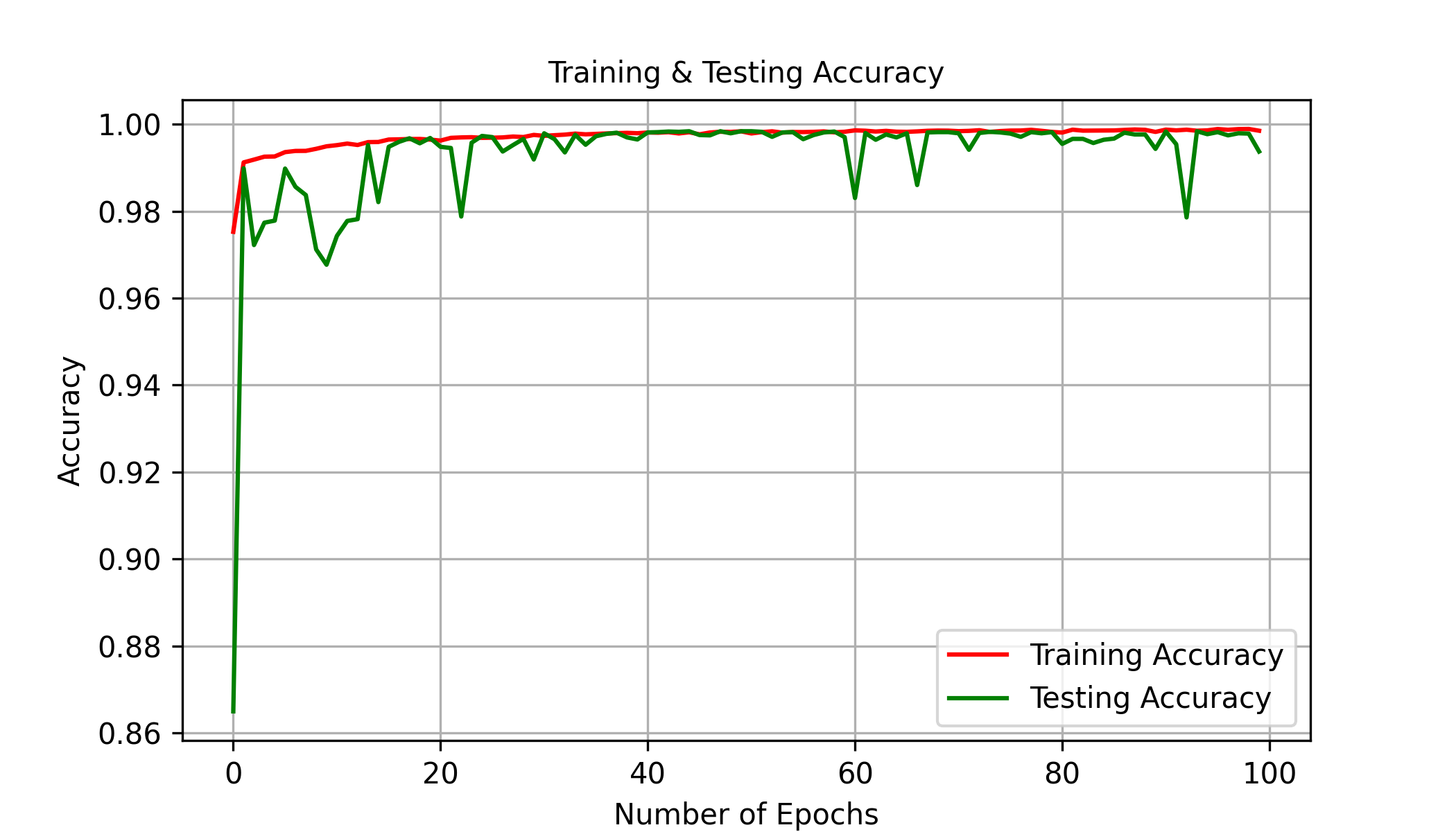}
    \caption{ }
    \label{fig:trainingTestingAccImg}
  \end{subfigure}
  \hspace{0.01\textwidth}
  \begin{subfigure}[b]{0.47\textwidth}
    \includegraphics[width=\textwidth]{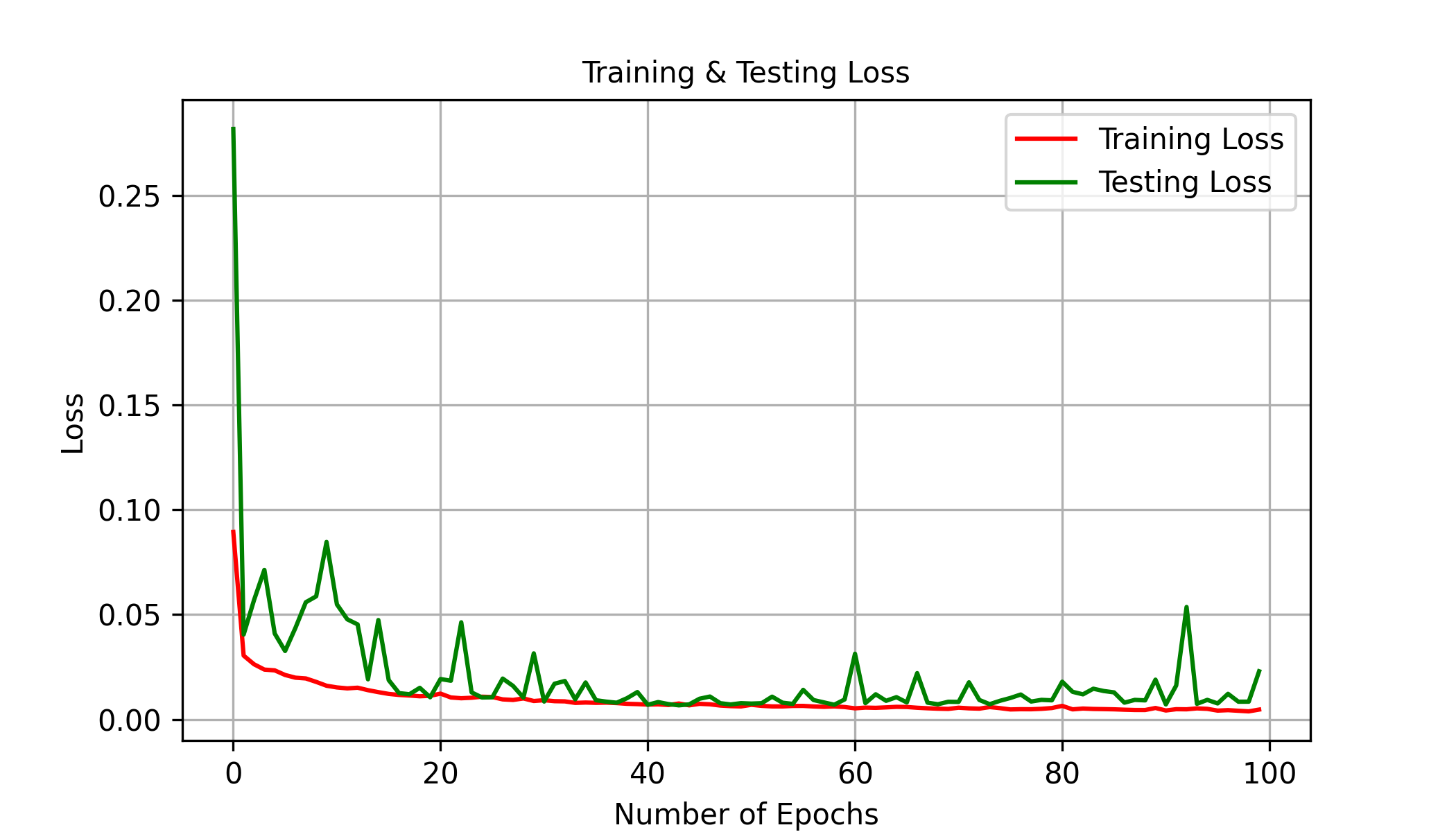}
    \caption{ }
    \label{fig:trainingTestingLossImg}
  \end{subfigure}
  
  \caption{(a) Training and Testing accuracy per epoch, (b) Training and Testing loss per epoch.}
  \label{fig:training_testing_acc_loss}
\end{figure}

Figures \ref{fig:trainingTestingAccImg} illustrated our proposed CNN model training and testing accuracy per epoch. This graph makes it quite evident that, in comparison to the testing accuracy score, the training accuracy score improved steadily. Testing accuracy was somewhat inconsistent over the first 25 epochs, later, it became consistent, and the model was successfully trained. Conversely, \ref{fig:trainingTestingLossImg} shows the proposed CNN model's training and testing loss score per epoch. While the training loss went down steadily, the testing loss fluctuated in between 5 to 20 epochs, and became consistent after 20 epochs. So, the model works well both for the training and testing phases.   

\subsection{Explainability Analysis}
\label{explainability}
SHAP is a useful method to understand how machine learning and deep learning models make predictions. It helps us to guess the importance of each feature in a prediction, explaining why the model decided on a certain outcome. In this research, we used the \textit{SHAP DeepExplainer} method to figure out the SHAP values for our CNN model. Since our dataset is extensive, we took 5000 samples to calculate these values.

\subsubsection{SHAP Global Interpretation}
\begin{figure}[!ht]
    \centering
    \includegraphics[width=0.5\textwidth, height=0.3\textheight]{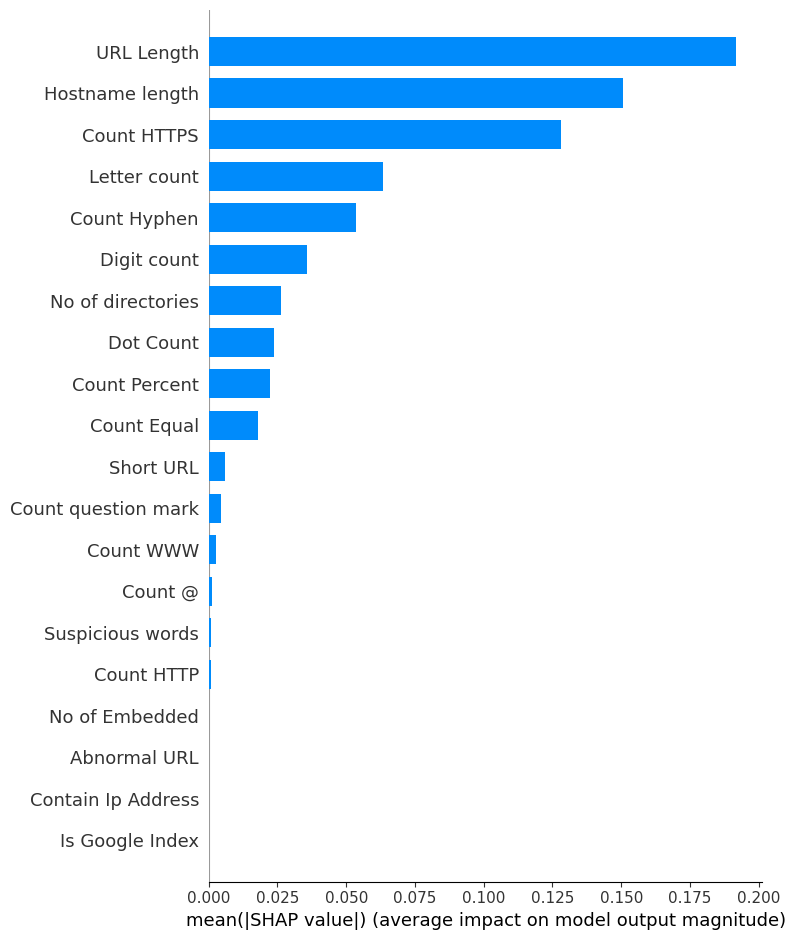}
    \caption{\centering Hierarchy of features with respect to their role in classification and Average impact on model output magnitude.}
    \label{fig:explainability1}
\end{figure}

The summary plot shows the most essential features and the magnitude of their impact on the model. Figure \ref{fig:explainability1} provides a hierarchical summary of the average effect on model output magnitude. 

The size of the SHAP values shows how much a feature affects the prediction. If a feature has bigger SHAP values (either positive or negative), it has a stronger effect on predicting phishing URLs. Notably, the feature "\emph{URL Length}" has the largest average impact on the model's classification, whereas "\emph{Count WWW}" contributes the least. Furthermore, "\emph{Hostname length}," is credited with having a major impact on phishing URL detection, contributing to the second-highest mean SHAP value. On the other hand, "\emph{Count question mark}" makes the second-smallest contribution to the classification. "\emph{Count @}," "\emph{Suspicious words}," "\emph{Count HTTP}," "\emph{No of Embedded}," "\emph{Abnormal URL}," "\emph{Contain Ip Address}" and "\emph{Is Google Index}" play least significant role to the classification of phishing URL detection. 


\subsubsection{SHAP Local Interpretation}
\begin{figure}[!ht]
    \centering
    \includegraphics[width=0.5\textwidth, height=0.3\textheight]{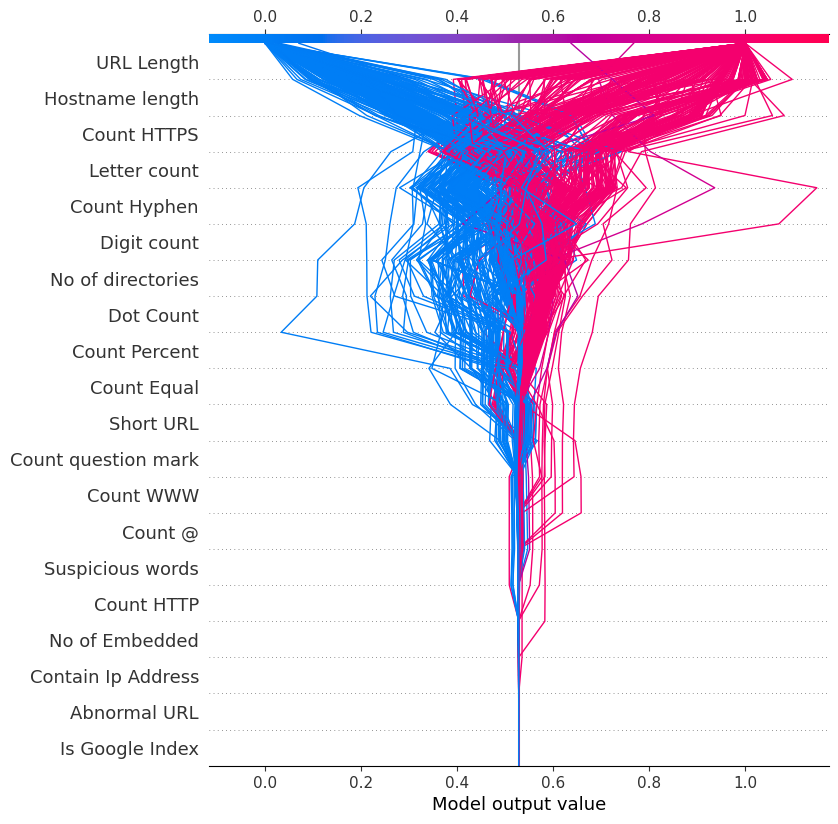}
    \caption{\centering SHAP value decision plot.}
    \label{fig:explainability3}
\end{figure}

\begin{figure}[!ht]
    \centering
    \includegraphics[width=0.5\textwidth, height=0.3\textheight]{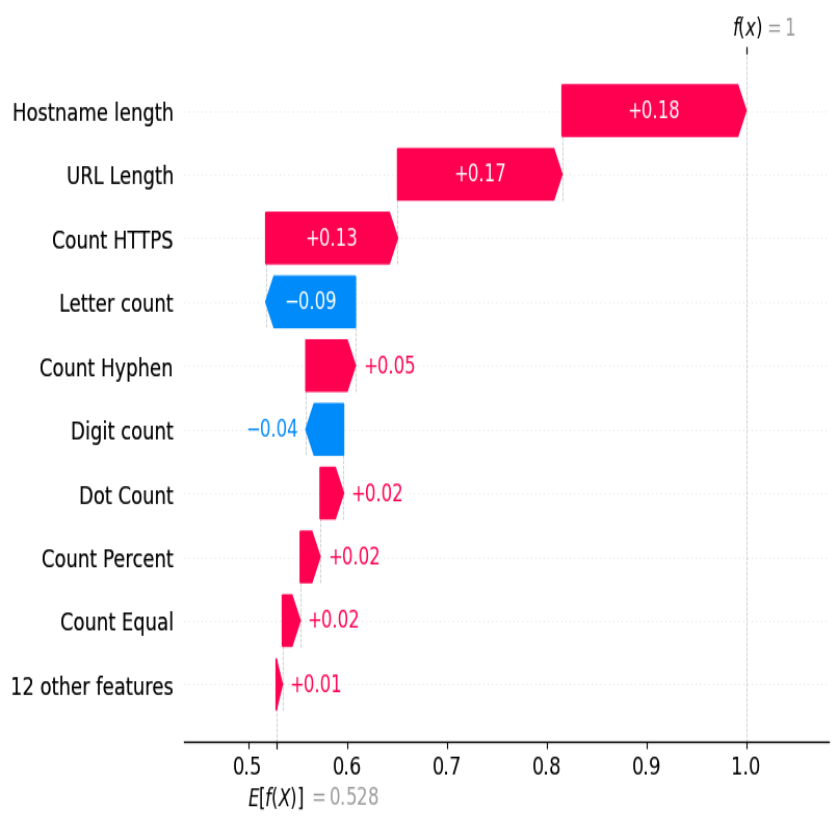}
    \caption{\centering SHAP value waterfall plot.}
    \label{fig:explainability4}
\end{figure}

Figure \ref{fig:explainability3}, shows the decision plot makes it possible to observe the amplitude of each change, taken by a sample for the values of the displayed features. In figure \ref{fig:explainability3}, each line shows what the model predicts for a specific example. The horizontal line in the middle represents the average prediction of the model. The features of the model are shown on the vertical axis. The colored lines start at the bottom and reach the middle line, representing the model's prediction for each example. The color of the line depends on the predicted value. As you move up the graph, we see how each feature adds to or subtracts from the model's prediction, showing the contribution of each feature to the overall prediction.
\newline
Figure \ref{fig:explainability4} shows the waterfall plot, which is a local analysis tool that uses a bar chart to display the positive and negative impacts of different features on a model's prediction. It starts with a baseline value and helps to identify the features that influence the model's predictions most. The plot shows the values of different features and their impact on the model's predictions. In figure \ref{fig:explainability4}, the features like \textit{Hostname length, URL Length} and other has a positive impact (shown in red), or the features like \textit{Letter count, Digit count} have a negative impact (shown in blue) on moving the value from what the model would expect based on the background data to the actual output for a specific prediction.

  


\subsection{Comparative Analysis}
It is important to compare our proposed model with other benchmark models to evaluate the acceptance of our model. We extracted the important features before jumping into the model training and it is one of our key contributions. The model has been compared both with the previous works on the same dataset \cite{PhishTank} and other ML models with our extracted features. 

\subsubsection{Performance comparison with conventional machine learning algorithms}
Table \ref{table:mlModelPerformance} compares the performance of several machine learning models with our proposed model on a classification task. As shown in the table, our proposed model achieves the highest accuracy (99.85\%) and F1-score (99.80\%) among all the models. This suggests that our model is the most effective at correctly classifying instances. The MultiLayerPerceptronClassifier model also performs well, with an accuracy of 99.78\% and an F1-score of 99.79\%. The other models have slightly lower accuracy and F1-score scores. These results demonstrate the effectiveness of our proposed model for this classification task.

\begin{table}[htbp]
\caption{Machine learning model performance comparison with the proposed model.}
\label{table:mlModelPerformance}
\begin{center}
\begin{tabular}{|c|c|c|c|c|}
\hline
 \textbf{Models} & \textbf{Accuracy} & \textbf{Precision} & \textbf{Recall} & \textbf{F1-Score}\\
 \cline{1-5}
    KNN & 99.47 & 99.79 & 99.20 & 99.49 \\
    DT & 99.67 & 99.71 & 99.65 & 99.68  \\
    RF & 99.12 & 99.10 & 98.88 & 98.98  \\
    XGB & 99.71 & 98.80 & 98.79 & 98.67 \\
    MLP & 99.78 & 99.72 & 99.86 & 99.79 \\
    \textbf{Our proposed model} & \textbf{99.85} & \textbf{99.90} & \textbf{99.80} & \textbf{99.80}\\

\hline
\multicolumn{5}{l}{}
\end{tabular}
\label{tab1}
\end{center}
\end{table}


\subsubsection{Performance comparison with previous works}
Table \ref{table:comparisonTable} compares the accuracy and explainability of our proposed model with other phishing URL detection models. The table demonstrates that, out of all the models, our 1D Convolutional Neural Network (CNN) model have the best accuracy (99.85\%). Along with the proposed model, 21 categorical and numerical features played a vital role in predicting higher accuracy.  Additionally, our model only provides explainability for its predictions. This means our model can identify phishing URLs accurately and explain why it classifies them as phishing. This is a significant advantage, as it can help users to understand how the model works and to trust its predictions. The other models listed in the table are all less accurate than ours, and none provide explainability. The Ozgur Koray Sahingoz et al. \cite{sahingoz2024dephides}, their CNN model is the second most precise model in the table and has an accuracy of 98.74\%, but it didn't explain their model.


\begin{table}[h]
\caption{Research comparison with others}\label{table:comparisonTable}
\centering
\begin{tabular}{|c|c|c|c|c|}
\hline
\textbf{Ref.} & \makecell{\textbf{Algorithm} \\ \textbf{Used}} & \makecell{\textbf{Features} \\ \textbf{Extraction}} & \textbf{Accuracy} &  \makecell{\textbf{Explain-} \\ \textbf{ability}} \\ [1.5ex]
\hline
\cite{korkmaz2022hybrid} & GCNN & Content-based & 98.37\% & No \\ [1.5ex]
\cite{ahammad2022phishing} & LightGBM & 12 features & 86.00\% & No \\ [1.5ex]
\cite{mehndiratta2023malicious} & KNN & 14 features & 90.00\% & No \\ [1.5ex]
\cite{sahingoz2024dephides} & CNN & URL Vectorization & 98.74\% & No \\ [1.5ex]
\makecell{\textbf{Our} \\ \textbf{Model}} & \textbf{1D CNN} & \textbf{21 features} & \textbf{99.85\%} & \textbf{Yes} \\ [1.5ex]
\hline
\end{tabular}
\end{table}

\section{Conclusion} \label{conclusion}
In this proposed phishing URL detection model, we have worked on creating an intelligent phishing website detection technique from URLs that might assist internet users. We have proposed a 1D Convolutional Neural Network (CNN) to detect the phishing URLs. As deep learning models are data-hungry, we have collected a huge amount of secondary data and carefully extracted 21 features from the URLs. It played a vital role in the creation of a robust model. After extensive experiments and comparisons, we have established that our model outperforms the existing models. Cybersecurity is a critical issue. So, the interpretability and explainability of the model are essential. To address the issue, we have shown the explainability of the model. It will help the user to be careful about the features that significantly contribute to the phishing URLs.
\section{Acknowledgements}
The authors would like to acknowledge that all authors contributed equally to this work.

\bibliographystyle{ieeetr}
\bibliography{bibliography}

\end{document}